\documentclass[prd,nofootinbib,showkeys,showpacs,twocolumn]{revtex4-1}

\usepackage{graphicx}
\usepackage{float}
\usepackage{amsmath}
\usepackage{cancel}
\usepackage{hyperref}
\usepackage{mathtools}
\usepackage{txfonts}

\newcommand{\trace}{\operatorname{tr}}
\newcommand{\te}{\text}
\newcommand{\nn}{\nonumber}
\newcommand{\cd}{\cdot}

\begin{document}

\title{Electromagnetic transition form factors of light vector mesons}
\author{Carla Terschl\"usen}
\affiliation{Institut f\"ur Theoretische Physik, Universit\"at Giessen, Germany}
\author{Stefan Leupold}
\affiliation{Institutionen f\"or fysik och astronomi, Uppsala Universitet, Sweden}

\begin{abstract}
The decays of narrow light vector mesons into pseudoscalar mesons and dileptons are calculated to leading order in a recently proposed scheme which treats pseudoscalar and vector mesons on equal footing. Since all required parameters have been determined by other reactions the presented approach gains predictive power for the considered processes. The decay of the $\omega$-meson into pion and dimuon agrees reasonably well with the available experimental data concerning form factor, single-differential decay width and partial decay width. As well do the partial decay width of the $\omega$-meson into a pion and a dielectron and of the $\phi$-meson into an $\eta$-meson and a dielectron. The decay properties of the $\omega$-meson into $\eta$-meson and dimuon or dielectron and of the $\phi$-meson into $\eta$-meson and dimuon are predicted.
\end{abstract}

\keywords{Vector mesons, electromagnetic form factors, vector-meson dominance, effective field theory}

\pacs{14.40.Be,13.40.Gp,12.40.Vv,11.30.Rd}

\maketitle


\section{Introduction}

The development of systematic approaches for the calculation of hadronic reactions and decays is one of the open problems of QCD. If energies are restricted to a region in which the only participating mesons are Goldstone bosons, chiral perturbation theory successfully describes the dynamics of the relevant degrees of freedom 
\cite{Weinberg:1978kz,Gasser:1983yg,Gasser:1984gg,Scherer:2002tk}. In the energy range of hadronic resonances, however, one
typically has phenomenologically successful models at hand, but, being models instead of effective field theories, it is not clear how to systematically improve them or assess quantitatively the intrinsic uncertainties. Clearly it is desirable to
push the borderline for the application of effective field theories towards higher energies. Recently a counting scheme
has been suggested in \cite{Lutz:2008km} and further explored in \cite{Leupold:2008bp} for flavor-SU(3) systems of 
Goldstone bosons and light vector mesons. Such a scheme makes in particular much sense if {\em all} degrees of freedom
relevant for the considered energy range are taken into account. Restricting the attention to pseudoscalar and vector mesons
is certainly reasonable if all other low-lying mesons can be understood as being dynamically generated from the interactions
of the former (concerning scalar and/or axial-vector mesons see, e.g., 
\cite{Oller:1998zr,Pelaez:2003dy,Lutz:2003fm,Roca:2005nm,Wagner:2008gz,Wagner:2007wy,Branz:2008ha} and references therein). 
This is essentially the hadrogenesis conjecture \cite{Lutz:2001dr,Lutz:2001yb,Lutz:2001mi,Lutz:2005ip,Lutz:2008km} applied to the sector of flavor-SU(3) mesons. 

In the counting scheme of \cite{Lutz:2008km} the masses of both vector mesons ($A$) and pseudoscalar mesons ($B$) are treated as soft, i.e.
\begin{align}
 m_A \sim Q  \,, \qquad m_B \sim Q
\label{eq:powercount1}
\end{align}
where $Q$ is a typical momentum. Focusing on decays of vector mesons, all involved momenta are necessarily smaller
than the vector-meson masses. Thus, a derivative always scales as 
\begin{align}
 \partial_\mu \sim Q
\label{eq:powercount2}
\end{align}
not depending on wether it acts on the vector or the pseudoscalar meson. This is conceptually different from the approach 
followed in \cite{Ecker:1988te,Ecker:1989yg}. While (part of) the Lagrangian of \cite{Lutz:2008km,Leupold:2008bp}, which we use in the following, resembles the one of \cite{Ecker:1988te,Ecker:1989yg}, there is an important difference in the power counting: In \cite{Ecker:1988te,Ecker:1989yg} all derivatives and the masses of pseudoscalar mesons are also treated as soft, but the vector-meson masses are not. Since our approach aims in the present work at the description of vector-meson decays, it is 
certainly suggestive to conceptually treat all masses, energies and momenta of the actively involved mesons on equal footing.
In \cite{Lutz:2008km} this scheme based on (\ref{eq:powercount1},\ref{eq:powercount2}) has been used in leading order to calculate two-body decays of the nonet of light vector mesons. It was possible to qualitatively explain the experimental finding 
that flavor breaking is rather small (of subleading order) for the corresponding hadronic and dilepton decays of vector mesons while it is sizable for the radiative decays into photon and pseudoscalar meson. Quantitatively these two-body decays have 
been used in 
\cite{Lutz:2008km} to fix the parameters (low-energy constants) of the leading-order Lagrangian. Extending this work,
hadronic three-body decays of vector mesons have been determined in \cite{Leupold:2008bp}. It turned out that no new parameters were needed for the leading-order calculation. The result for the decay width of the $\omega$-meson into three pions 
turned out to be very close to the experimental value. 
Predictions for rare $K^*$-meson decays into two pions and one kaon have been presented.
It is the purpose of the present work to extend this study to electromagnetic transition form factors. Again, no new parameters
are needed for the leading-order calculation. Therefore, it is possible to test the results against the available data 
and to predict decay rates for processes which were not measured yet.

Electromagnetic form factors are regarded as an important tool to study the intrinsic structure of 
hadrons \cite{Landsberg:1986fd}. Of particular interest are the decays of the narrow pseudoscalar states $\pi$,
$\eta$ and $\eta'$ into
two real or virtual photons and of the narrow vector states $\omega$ and $\phi$ into a pseudoscalar and a (real or
virtual) photon. The pseudoscalar decays are even of some relevance for searches for physics beyond the
standard model of elementary particles: Their size influences the anomalous magnetic moment of the 
muon \cite{Bijnens:2007pz}. On the other hand, the mentioned decays of pseudoscalar mesons and of vector mesons 
are even interrelated, since the neutral vector mesons have the same quantum numbers as the photon.
In fact, for many reactions, where dileptons couple (via a virtual photon) to hadrons, 
the vector-meson dominance (VMD) assumption \cite{sakurai-book} turned out to be phenomenologically very successful. One example 
where standard VMD dramatically fails, however, is the transition form factor of the 
$\omega$-meson \cite{Landsberg:1986fd,:2009wb}, i.e.\ one of the quantities which we will study here. 
In fact, previous models based on VMD, e.g., \cite{Klingl:1996by,Faessler:1999de}, could not explain the 
steep rise of the $\omega$ form factor.
On the other hand, our counting scheme provides a new microscopic view on VMD and systematic corrections to it: 
Since in the VMD contribution to a process a vector-meson propagator appears, such a contribution is typically enhanced relative 
to the corresponding point interaction by two orders in the scale $Q$ of typical momenta. For the case of interest this will
be shown explicitly below in Sect.\ \ref{sec:basics}. Of course, there are corrections to that picture which come in at
next-to-leading order of our counting scheme. To work out these corrections is left for future work. In the following we
restrict ourselves to the leading-order calculation. 
Yet there is still a twist in the argument concerning the enhancement of the VMD contribution. 
It comes from a technical aspect of our formalism: 
Our counting scheme is formulated for vector mesons in the tensor realization using six degrees of freedom with three frozen
to describe the three physical spin states of a massive vector meson. In contrast, standard VMD is formulated
in the vector realization which uses four degrees of freedom (with one frozen). It turns out that in the tensor realization
the leading-order contribution to the transition form factors indeed comes solely from diagrams with intermediate vector mesons.
However, if one translates the contribution into the language of standard VMD one gets both a vector-meson contribution and
a contact term. The crucial point is that no new parameters show up here. The size of the contact term is fixed by the
fact that it originates from the vector-meson contribution in the tensor realization. This and only this contact term is
of the same order in our counting scheme as the (standard) vector-meson contribution. All these issues will be further
discussed below. Concerning the different realizations of vector-meson fields and their interrelations we refer to
\cite{Ecker:1989yg,Bijnens:1995ii,Leupold:2006bp}.
We note in passing that the same finding holds for the three-pion decay of the $\omega$-meson studied
in \cite{Leupold:2008bp}. 

In the following we will study the processes $\omega \rightarrow  \pi^0 \, l^+ l^-$, 
$\omega \rightarrow \eta \, l^+ l^-$ and $\phi \rightarrow \eta \, l^+ l^-$ with leptons $l=e,\mu$. The OZI forbidden decay 
$\phi \rightarrow \pi^0 \, l^+ l^-$ is not covered by the leading-order Lagrangian of our scheme 
(cf.\ also \cite{Lutz:2008km}).
For the decay $\omega \rightarrow \pi^0 \, \mu^+ \mu^-$ very recent and accurate experimental data for the 
form factor are available, provided by the NA60 collaboration \cite{:2009wb}. 
In addition, experimental values for the (integrated) partial decay width of the $\omega$-meson into a pion and both dielectron and dimuon are collected in \cite{Amsler:2008zzb}. There are also data taken with the SND detector at the VEPP-2M 
collider for the transition form factor of the $\phi$-meson to an $\eta$-meson and a dielectron \cite{Achasov:2000ne}. Finally, the experimental value for the corresponding partial decay width can be taken from \cite{Amsler:2008zzb}. Besides these comparisons to existing data we provide 
predictions for the decays of the $\omega$-meson into an $\eta$-meson and a dielectron or dimuon and for the decay of 
the $\phi$-meson into an $\eta$-meson and a dimuon.

The paper is organized in the following way: In the next section the relevant part of our leading-order Lagrangian and 
general formulae for the transition form factor are introduced. 
The calculations for the decay of an $\omega$-meson into a pion and a dilepton are presented in Sect.\ \ref{sec:omega pi}. For all considered processes the form factor, the single-differential decay width and the integrated partial decay width as well as the corresponding branching ratio are given. Additionally, the single-differential decay width of decays into dielectrons is integrated for dielectron masses above $2m_\mu$ to be able to compare results of the decays into dielectrons and dimuons. In Sects.\ \ref{sec:omega eta} and \ref{sec:phi eta} we present the results for the decay of an $\omega$-meson and 
a $\phi$-meson, respectively, into an $\eta$-meson and a dilepton. 
Finally, the results will be summarized and an outlook on possible extensions of the present work
will be given in Sect.\ \ref{sec:summary}.


\section{Transition form factor and leading-order Lagrangian} 
\label{sec:basics}

Generically the matrix element for the decay of a vector meson $A$ into a dilepton $l^+ l^-$ and a pseudoscalar meson $B$ can
be expressed as \cite{Landsberg:1986fd}
\begin{align}
\mathcal{M}(A \rightarrow B\,l^+l^-) = e^2 \, f_{AB}(q^2) \, \epsilon^{\mu\nu\alpha\beta} q_\mu k_\nu \epsilon_\alpha \, 
\frac{1}{q^2} \, \bar{u}_s(q_1) \gamma_\beta v_{s'}(q_2)  \,.
\end{align}
Here $e$ is the electron charge, 
$\epsilon^{\mu\nu\alpha\beta}$ denotes the Levi-Civita tensor, $k$ and $q$ are the four-momenta of pseudoscalar meson 
$B$ and the virtual photon, respectively, $\epsilon_\alpha$ is the polarization four-vector of the vector meson $A$, 
$q_{1,2}$ is the four-momentum of the lepton $l^{\mp}$ and $u$, $v$ denote the corresponding spinors. 
The hadronic information is included in the form factor $f_{AB}(q^2)$ of the $A \rightarrow B$ transition.
It is common practice to introduce a normalized form factor as  
\begin{align} 
F_{AB}(q^2) := \frac{f_{AB}(q^2)}{f_{AB}(0)} \,, 
\label{eq:ff-norm}
\end{align}
so that $F_{AB}(0)=1$ at the photon point. 

The double-differential decay rate of the decay of a vector meson $A$ into a pseudoscalar meson $B$ and a dilepton $l^+ l^-$ can be calculated as \cite{Amsler:2008zzb} 
\begin{align}
 \frac{\te{d}^2 \Gamma_{A \rightarrow Bl^+l^-}}{\te{d}m_{l^+l^-}^2 \te{d}m_{l^+B}^2} = \frac{2\alpha^2}{\pi} \frac{1}{32 m_A^3} |f_{AB}(q^2)|^2 \frac{P}{q^4}
 \label{eq:double-diff decay}
\end{align}
with the fine-structure constant $\alpha=e^2/(4\pi)$, the phase-space factor
\begin{eqnarray} 
 P & = & - \frac{1}{3} \, \epsilon^{\mu\nu\alpha\beta} k_\mu q_\nu \, 
\epsilon_{\bar{\mu}\bar{\nu}\alpha \bar{\beta}} \, k^{\bar{\mu}} q^{\bar{\nu}}
\nonumber \\ && \times 
\sum_{s,s'} \bar{u}_s(q_1) \gamma_\beta v_{s'}(q_2) \, \bar{v}_{s'}(q_2) \gamma^{\bar{\beta}} u_s(q_1) 
\end{eqnarray}
and the following variables:
\begin{align} 
m_{l^+l^-}^2 := (q_1 + q_2)^2 = q^2 \,, \quad m_{l^+B}^2:=(q_2 + k)^2  \,. 
\end{align}
By integrating Eq.\ \eqref{eq:double-diff decay} the single-differential decay width \cite{Landsberg:1986fd}
\begin{align}
 \frac{\te{d} \Gamma_{A \rightarrow B \, l^+ l^-}}{\te{d}m_{l^+l^-}^2 \ \Gamma_{A \rightarrow B \gamma}} = \frac{\alpha}{3\pi} \sqrt{1-\frac{4 m_l^2}{m_{l^+l^-}^2}} \, \left( 1+\frac{2 m_l^2}{m_{l^+l^-}^2} \right) \, \frac{1}{m_{l^+l^-}^2} 
\nonumber \\ \times
 \left[ \left(1+\frac{m_{l^+l^-}^2}{m_A^2 - m_B^2} \right)^2 - \frac{4 m_A^2 m_{l^+l^-}^2}{(m_A^2-m_B^2)^2} \right]^{3/2} \, 
|F_{AB}(m_{l^+l^-}^2)|^2 
\label{eq:single-diff}
\end{align}
is obtained. Here $m_l$ is the lepton mass and 
\begin{align}
 \Gamma_{A \rightarrow B \gamma} = \frac{\left(m_A^2 - m_B^2 \right)^3 e^2}{96 \pi \, m_A^3} \, |f_{AB}(0)|^2 
\label{eq:BreiteAgamma}
\end{align}
denotes the partial decay width into a real photon. 

The task is to obtain an expression for the hadronic quantity $f_{AB}(q^2)$. In fact, our formalism ``only''
provides a prediction
for the normalized form factor \eqref{eq:ff-norm} 
since our input parameters have been fitted in \cite{Lutz:2008km} to obtain the experimental
values for the real-photon radiative decay widths \eqref{eq:BreiteAgamma} for various combinations $A$ and $B$.
Since the experimental values for the latter decay widths have slightly changed and to obtain a rough estimate for the intrinsic  
uncertainties of our approach based on a leading-order calculation we will provide updated fits of our parameters below.

In principle, the vector meson $A$ can either directly decay into $B$ and a (real or virtual) photon or indirectly
via an intermediate vector meson. 
The leading-order Lagrangian of \cite{Lutz:2008km} allows only for the indirect decay. Its relevant part is given by
\begin{align}
\mathcal{L}_{\rm indir.} = & -\frac{h_A}{16f} \epsilon^{\mu\nu\alpha\beta} 
\trace \{[V_{\mu\nu},(\partial^\tau V_{\tau\alpha})]_+  \, \partial_\beta \Phi\} 
\nonumber \\  
                   & - \frac{b_A}{16f} \epsilon^{\mu\nu\alpha\beta} \trace \{[V_{\mu\nu},V_{\alpha\beta}]_+ \, [\Phi,\chi_0]_+\} 
\nonumber \\ 
		   & - \frac{e_V m_V}{4} \trace \{V^{\mu\nu} Q \, \partial_\mu A_\nu  \}   \,.
\label{eq:Lid}
\end{align}
The first two terms of this Lagrangian describe the decay of the vector meson $A$ into the virtual meson $A'$ and the pseudoscalar meson $B$ while the last term yields the direct conversion of the meson $A'$ into a photon. (The final decay of the virtual photon into the lepton pair is described by usual QED.)
In \eqref{eq:Lid} the following flavor matrices appear: The vector and pseudoscalar mesons are collected in
\begin{align}
V_{\mu\nu} = \begin{pmatrix}
                 \rho^0_{\mu\nu} + \omega_{\mu\nu} & \sqrt{2}\rho^+_{\mu\nu} & \sqrt{2}K^+_{\mu\nu} \\
                 \sqrt{2}\rho^-_{\mu\nu} & -\rho^0_{\mu\nu}+\omega_{\mu\nu} & \sqrt{2}K^0_{\mu\nu} \\
                 \sqrt{2}K^-_{\mu\nu} & \sqrt{2}\bar{K}^0_{\mu\nu} & \sqrt{2}\phi_{\mu\nu}
              \end{pmatrix}
\end{align}
and
\begin{align}
\Phi = \begin{pmatrix}
           \pi^0+\frac{1}{\sqrt{3}}\eta & \sqrt{2}\pi^+ & \sqrt{2}K^+ \\
	   \sqrt{2}\pi^- & -\pi^0+\frac{1}{\sqrt{3}}\eta & \sqrt{2}K^0 \\
           \sqrt{2}K^- & \sqrt{2}\bar{K}^0 & -\frac{2}{\sqrt{3}}\eta
        \end{pmatrix} \,,
\end{align}
respectively. The quark charge matrix is denoted by
\begin{align}
Q &= \begin{pmatrix}
        \frac{2}{3} & 0 & 0 \\
	0 & -\frac{1}{3} & 0 \\
 	0 & 0 & -\frac{1}{3}
     \end{pmatrix}
\end{align}
and the flavor-breaking term proportional to the quark mass matrix is
\begin{align}
\chi_0 &= \begin{pmatrix}
            m_\pi^2 & 0 & 0 \\
            0 & m_\pi^2 & 0 \\
 	    0 & 0 & 2m_K^2-m_\pi^2
          \end{pmatrix}
\end{align}
with the pion mass $m_\pi$ and the kaon mass $m_K$ neglecting isospin-breaking effects. Finally $A_\nu$ denotes the photon field.
The values for the coupling constants $h_A$, $b_A$, $f$, $e_V$ and $m_V$ will be given below.

Since in our approach large-$N_c$ considerations are incorporated \cite{Lutz:2008km}, loops are automatically suppressed and 
therefore do not show up in a leading-order calculation. Hence the calculation is restricted to the tree-level contributions 
emerging from \eqref{eq:Lid}. In \cite{Lutz:2008km} one particular tree-level next-to-leading-order term has been selected to obtain
a rough estimate about the importance of next-to-leading-order contributions. We also provide such estimates here for the
transition form factors. To this end we introduce one part of the next-to-leading-order Lagrangian:
\begin{align}
\mathcal{L}_{\rm dir.} = -\frac{e_A}{4fm_V} \epsilon^{\mu\nu\alpha\beta} \trace \{[Q,(\partial^\tau V_{\tau\alpha})]_+ 
\, \partial_\beta \Phi \, \partial_\mu A_\nu \}   \,.
\label{eq:Ld}
\end{align}
It is important to stress that we do not provide a full next-to-leading-order calculation here. We even do not provide the
(relevant part of the) complete next-to-leading-order Lagrangian. The term given in \eqref{eq:Ld} is only a selection.
However, besides the possibility to provide rough error estimates, this particular term serves to further discuss the issue of
vector-meson dominance (VMD). 
Indeed, the term given in \eqref{eq:Ld} describes the direct decay of $A$ into $B$ and a photon without an intermediate
vector meson, i.e.\ it is a non-VMD term. We will come back to that point in Sect.\ \ref{sec:omega pi} where we discuss the
form factor of the $\omega$-meson. 
In the remainder of the present section we will point out how the coupling constants appearing in \eqref{eq:Lid} and \eqref{eq:Ld}
are fixed. 

Following \cite{Lutz:2008km} and previous works cited therein we choose $f\,$$\simeq$$\,90\,$MeV for the pion decay constant in the 
three-flavor chiral limit. The scale $m_V\,$=$\,776\,$MeV has been introduced for convenience to obtain dimensionless coupling
constants $e_V$ and $e_A$. The parameter $e_V\,$$\simeq$$\,0.22$ is fixed by the direct dilepton decays of $\rho^0$, $\omega$ and $\phi$
\cite{Lutz:2008km}. The remaining parameters are now fitted to the decays $\omega\to\pi^0 \gamma$, $\omega \to \eta \gamma$
and $\phi\to \eta\gamma$, i.e.\ to the real-photon counterparts of the transition form factors we are interested in. 
First we use a strict leading-order set, i.e.\ we put $e_A\,$=$\,0$ and choose
\begin{align}\tag{P1} 
h_A \simeq 2.32 \,, \qquad b_A \simeq 0.19  \,. 
\label{Ps1}
\end{align}
The good quality of the fit is demonstrated in Table \ref{tab:Parameter}. A second parameter set is obtained by allowing for
a non-vanishing value of $e_A$. This procedure has already been performed in \cite{Lutz:2008km} (considering in addition also
radiative $K^*$ decays). Meanwhile the data for these decays have slightly changed \cite{Amsler:2008zzb}. We take the values for the leading-order parameters $h_A$ and $b_A$ from \cite{Lutz:2008km} and only fine-tune $e_A$. In that way we get
\begin{align}
\tag{P2} 
h_A \simeq 2.10  \,, \quad b_A \simeq 0.27 \,, \quad e_A \simeq 0.015  \,. 
\label{Ps2} 
\end{align}
We did not modify the values for $h_A$ and $b_A$ from \cite{Lutz:2008km} since they gave an excellent description for the 
three-pion decay of the $\omega$-meson \cite{Leupold:2008bp}. We note that the value for $e_A$ is only marginally changed: It
was $e_A\,$$\simeq$$\,0.02$ in \cite{Lutz:2008km}. Also for parameter set \eqref{Ps2} the
resulting values for the radiative decay widths are given in Table \ref{tab:Parameter}. 
\begin{table}[h]
\begin{tabular}{l|l|l|l}
 & exp. value & param.\ set \eqref{Ps1} & param.\ set \eqref{Ps2} \\ \hline
$\Gamma_{\omega \rightarrow \pi^0 \gamma}$ & $(7.03 \pm 0.30) \cd 10^{-4} \,\te{GeV}$ & $7.14 \cd 10^{-4} \,\te{GeV}$ & $7.34 \cd 10^{-4} \,\te{GeV}$ \\ 
$\Gamma_{\omega \rightarrow \eta \gamma}$ & $(3.91 \pm 0.38) \cd 10^{-6} \,\te{GeV}$ & $3.71 \cd 10^{-6} \,\te{GeV}$ & $3.83 \cd 10^{-6} \,\te{GeV}$ \\
$\Gamma_{\phi \rightarrow \eta \gamma}$ & $(5.58 \pm 0.15) \cd 10^{-5} \,\te{GeV}$ & $5.38 \cd 10^{-5} \,\te{GeV}$ & $5.12 \cd 10^{-5} \,\te{GeV}$
\end{tabular}
\caption{Partial decay width calculated with parameter set \eqref{Ps1} and parameter set \eqref{Ps2}, respectively, compared to the experimental values as collected in \cite{Amsler:2008zzb}.}
\label{tab:Parameter}
\end{table}

In the following we will present calculations for both parameter sets \eqref{Ps1} and \eqref{Ps2}. As already mentioned
the purpose is to obtain a rough estimate for the inherent uncertainties of the approach caused by the fact that only a 
leading-order calculation is performed. We note that the values for the leading-order coupling constants $h_A$ and $b_A$
do not drastically differ for the two parameter sets. This gives credit to the proposition that the $e_A$-term of \eqref{eq:Ld} 
is of subleading order. In the following sections we study the transition form factors and the corresponding differential
decay widths. We stress again that for these calculations no new parameters appear.


\section{Decay $\omega \rightarrow \pi^0 \, l^+ l^-$} 
\label{sec:omega pi}

Analyzing the Lagrangian \eqref{eq:Lid} for the indirect decay of an $\omega$-meson into a pion and a vector meson, it turns out 
that due to isospin symmetry all terms vanish except the ones including $\rho$-mesons. 
So an $\omega$-meson can only decay into a pion and a dilepton via a virtual $\rho$-meson.
The form factor for direct plus indirect decay determined by the Lagrangians \eqref{eq:Ld} and \eqref{eq:Lid} is calculated as
\begin{align}
 f_{\omega\pi^0}(q^2) =& \frac{m_\omega}{2f \, m_V \, e} 
\left[ e_A  + 2 \, b_A \, e_V \, m_V^2 \, \frac{m_\pi^2}{m_\omega^2} \, S_\rho(q^2)\right. 
\nonumber \\
 & \left. - \frac{1}{4} \, e_V \, h_A \, m_V^2 \, \left(1+\frac{q^2}{m_\omega^2} \right) \, S_\rho(q^2) \right]
\label{eq:oursompi}
\end{align}
with the $\rho$-meson propagator \cite{Leupold:2008bp}
\begin{align}
 S_\rho(q^2) = \frac{1}{q^2 - m_\rho^2 + i\sqrt{q^2} \, \Gamma_\rho(q^2)}. 
\label{eq:rho propagator}
\end{align}
Here we have included for completeness the energy-dependent width,
\begin{align}
 \Gamma_\rho(q^2) = \Gamma_0 \, \left[\frac{p_{cm}(q^2)}{p_{cm}(m_\rho^2)} \right]^3 \frac{m_\rho^2}{q^2} \,,
\end{align}
of the $\rho$-meson with its onshell width $\Gamma_0\,$$\simeq$$\,150\,$MeV 
and the center-of-mass momentum of the pions
\begin{align}
 p_{cm}(q^2) = \frac{1}{2} \sqrt{q^2-4m_\pi^2}  \,.
\end{align}
It turned out that in all calculations the results with and without vector-meson widths differ by less than $1\%$
for the integrated quantities. As we do not consider our leading-order calculations and the determination of our parameters to have such a good accuracy, the width is neglected in the calculations presented in this paper. 
In addition, we note that for the unintegrated quantities (form factors) the modifications caused by a finite width 
are of the same size as the deviations between our two parameter sets. This is intrinsically consistent since
the finite width is indeed a next-to-leading-order effect. 

\begin{figure}[h]
 \centering
 \includegraphics[keepaspectratio,width=0.35\textwidth]{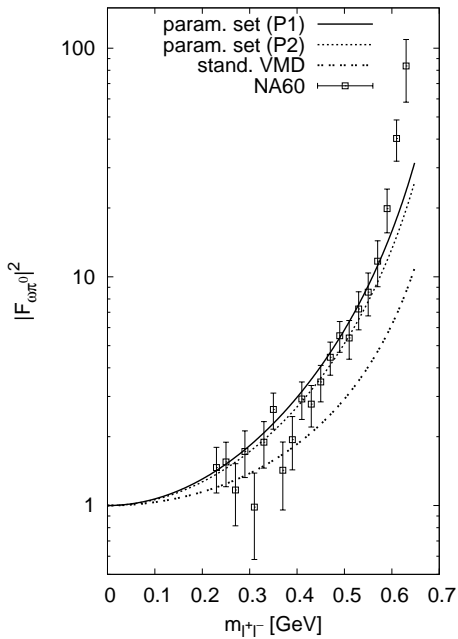}
 \caption{Form factor of the decay $\omega \rightarrow \pi^0 \, l^+ l^-$ compared to dimuon data taken by the NA60 
collaboration \cite{:2009wb}. The solid line describes the form factor calculated with parameter set \eqref{Ps1} and the 
dotted line the one calculated with parameter set \eqref{Ps2}. The dot-dashed line is calculated with the VMD model 
\eqref{eq:standardVMD} using the mass of the $\rho$-meson, $m_{\te{virtual}}=m_\rho$.}
 \label{fig:Ff omega pi}
\end{figure}

In Fig.\ \ref{fig:Ff omega pi} the respective form factor \eqref{eq:ff-norm} 
for both parameter sets \eqref{Ps1} and \eqref{Ps2} is plotted. The deviation of these two curves (full and dotted) from each
other is rather small, suggesting that the leading-order calculation could be reasonably accurate. 
Our calculations are compared on the one hand to the form factor which one gets from the 
standard vector meson dominance (VMD) assumption (dot-dashed line), i.e. 
\begin{align}
 F_{\te{VMD}}(q^2) = \frac{m_{\te{virtual}}^2}{m_{\te{virtual}}^2-q^2}
\label{eq:standardVMD}
\end{align}
with the mass $m_{\te{virtual}}$ of the intermediate vector meson, and on the other hand to the experimental data for the 
decay $\omega \rightarrow \pi^0 \, \mu^+ \mu^-$ obtained by the NA60 collaboration \cite{:2009wb}. 
Obviously, the VMD model fails to describe the data, the deficiency mentioned already in the introduction.
On the other hand, our calculations fit quite well to the experimental data, 
except for the last three data points which are already close to the upper kinematical boundary, 
$m_{l^+ l^-} \le m_\omega - m_\pi $. 

Before we continue with a comparison to further data it is worth to discuss qualitatively the difference of our approach
to standard VMD. Looking at the non-normalized form factor \eqref{eq:oursompi} we see that the $e_A$ term is clearly
a non-VMD term, a constant, while the $b_A$ term is of VMD type. The $h_A$ term, however, is of mixed character:
Neglecting the width in the vector-meson propagator one can rewrite 
\begin{align}
q^2 \, S_\rho(q^2)  \approx 1 + m_\rho^2 \,  S_\rho(q^2) \,.
\end{align}
Thus, this contribution consists of a constant (non-VMD term) and a term with a propagator (VMD term). Both are parts of
the $h_A$ contribution. On the other hand, numerically the $h_A$ term is the most important one. If one drops all other terms,
$e_A, b_A \to 0$, and neglects the difference between $\rho$- and $\omega$-meson masses, 
$m_\omega \to m_\rho$, one obtains from \eqref{eq:oursompi} for the normalized form factor \eqref{eq:ff-norm}
\begin{align}
F_{\omega\pi^0}(q^2) \to  \frac{m_\rho^2 + q^2}{m_\rho^2 - q^2} 
\end{align}
which should be compared to the standard-VMD formula \eqref{eq:standardVMD}.
Therefore, in the language of standard VMD our approach predicts a sizable deviation from VMD. It is important to stress,
however, that the additional non-VMD term (contact term) does not show up with an arbitrary adjustable parameter. Instead,
both the standard-VMD term and the contact term emerge from one and the same vector-meson contribution 
in our framework where vector mesons are represented by antisymmetric tensor fields. 
Though the chosen representation does not influence the resulting physics, contact terms can look differently in various representations and thus yield different orders in the applied counting scheme.
Hence, the deeper question is,
why one should use antisymmetric tensor fields together with the counting scheme and not any other representation. 
At present we cannot give a fully convincing answer to this question. Maybe this can only be provided by a microscopic
justification of our approach, i.e.\ by a determination of our low-energy constants from QCD. This is clearly far beyond
the scope of the present work. 
It is interesting to note, however, that among all representations which treat the vector mesons as transforming like 
ordinary matter fields under chiral transformations (and not like gauge fields) \cite{Ecker:1989yg}
it is the tensor representation where the terms with one vector-meson field have the minimal number of 
derivatives.\footnote{In the tensor representation, such terms contribute at $o(Q^2)$, in the vector representation
at $o(Q^3)$. In any other representation vector-meson fields have more than two indices and therefore involve 
more than two derivatives.} Therefore,
the vector-meson contributions are maximally enhanced. In that sense it can be reasonable to actually formulate the VMD
concept in the tensor representation; an observation also made in \cite{Ecker:1988te,Ecker:1989yg} in the context of
saturating the low-energy constants of pure chiral perturbation theory by vector mesons.
We note in addition that one motivation to introduce the tensor representation is the fact that current conservation
is easy to ensure, even in the presence of resummations \cite{Lutz:2007sk,Leupold:2006bp}.
In any case, from the comparisons to experimental data in the present work and in \cite{Leupold:2008bp} we conclude that
the use of the tensor representation together with our counting scheme provides reasonable results. Therefore we regard it as
worth-while to explore further consequences of our scheme in the future. 

\begin{figure}[h]
 \centering
 \includegraphics[angle=-90,keepaspectratio,width=0.45\textwidth]{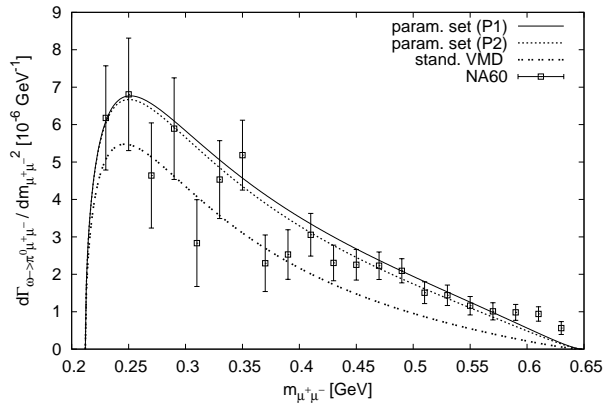}
 \caption{Single-differential decay width of the decay $\omega \rightarrow \pi^0 \, \mu^+ \mu^-$ compared to experimental data calculated with \eqref{eq:single-diff} and the form factor determined by the NA60 collaboration \cite{:2009wb}. 
The solid/dotted line 
describes the width calculated with parameter set \eqref{Ps1}/\eqref{Ps2}. The dot-dashed line is calculated with the VMD model.}
 \label{fig:single-diff decay omega pi mu}
\end{figure}

We continue our presentation by comparing our calculations to the 
single-differential decay widths given in \eqref{eq:single-diff}.
For dimuons the results are shown in 
Fig.\ \ref{fig:single-diff decay omega pi mu}. It is worth to point out how this figure is obtained. For the curves of
our approach (full and dotted line) we directly use the integral of \eqref{eq:double-diff decay} 
together with \eqref{eq:oursompi} and either \eqref{Ps1} or \eqref{Ps2}. For the VMD model (dot-dashed line) 
and to translate the form-factor
data of NA60 we used \eqref{eq:single-diff} together with the experimental value for $\Gamma_{\omega \to \pi^0 \gamma}$
(see Table \ref{tab:Parameter}). As can be seen in Fig.\ \ref{fig:single-diff decay omega pi mu} 
the high-mass data points which cannot be described by our approach (cf.\ Fig.\ \ref{fig:Ff omega pi})
will not contribute much to the integrated partial decay width. 

\begin{figure}[h]
 \centering
 \includegraphics[angle=-90,keepaspectratio,width=0.45\textwidth]{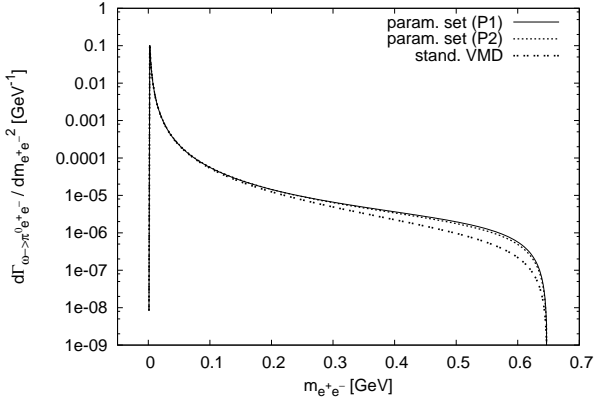}
 \caption{Single-differential decay width of the decay $\omega \rightarrow \pi^0 \, e^+ e^-$ calculated with
both parameter sets \eqref{Ps1} (solid line) and \eqref{Ps2} (dotted line) and the standard VMD form factor (dot-dashed line).}
 \label{fig:single-diff decay omega pi e}
\end{figure}

\begin{figure}[h]
 \centering
 \includegraphics[angle=-90,keepaspectratio,width=0.45\textwidth]{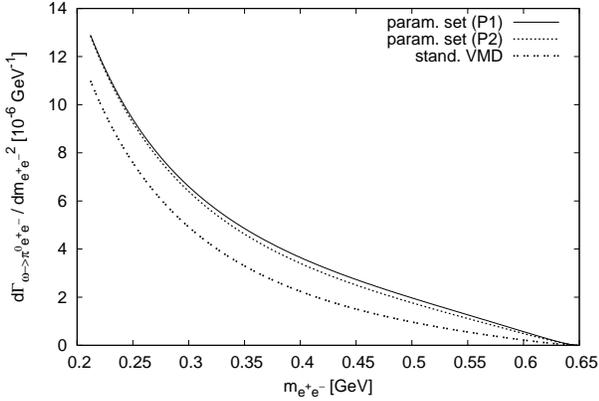}
 \caption{Single-differential decay width of the decay $\omega \rightarrow \pi^0 \, e^+ e^-$ above $m_{e^+e^-} = 2m_\mu$. Again, the solid/dotted line is calculated with parameter set \eqref{Ps1}/\eqref{Ps2} and the dot-dashed line with the VMD form factor.}
 \label{fig:single-diff decay omega pi e teil}
\end{figure}

The single-differential decay width of the $\omega$-meson into a pion and a dielectron is plotted in 
Fig.\ \ref{fig:single-diff decay omega pi e}. Obviously the peak appears at dielectron masses where the form factor
is hardly probed. Deviations between different form factors appear in the tail of the distribution. 
To be able to compare the results of the decay of the $\omega$-meson into a pion and a dielectron to those of the decay into a dimuon, the single-differential decay width above $m_{e^+e^-} = 2 m_\mu$ is plotted in Fig.\ \ref{fig:single-diff decay omega pi e teil}. Again one observes significant differences between our approach and standard VMD.

For the partial decay widths one gets
\begin{align}
 \Gamma_{\omega \rightarrow \pi^0 \mu^+ \mu^-} &= (9.85 \pm 0.58) \cd 10^{-7} \,\te{GeV}, \\
 \Gamma_{\omega \rightarrow \pi^0 e^+ e^-} &= (6.93 \pm 0.09) \cd 10^{-6} \,\te{GeV}
\end{align}
which agree very well with the experimental values from \cite{Amsler:2008zzb},
\begin{align}
 \Gamma_{\omega \rightarrow \pi^0 \mu^+ \mu^-}^\te{ exp} &= (8.15 \pm 2.13) \cd 10^{-7} \,\te{GeV}, \\
 \Gamma_{\omega \rightarrow \pi^0 e^+ e^-}^\te{ exp} &= (6.54 \pm 0.54) \cd 10^{-6} \,\te{GeV}.
\end{align}
Note that the value provided by NA60 \cite{:2009wb} for $\Gamma_{\omega \rightarrow \pi^0 \mu^+ \mu^-}$ is 
$(14.7 \pm 3.3) \cd 10^{-7} \,$GeV,
i.e.\ somewhat larger than the one given
in \cite{Amsler:2008zzb}, but compatible to it within two standard deviations.\\
Taking the full width of the $\omega$-meson $\Gamma_\omega = (8.49 \pm 0.08) \ \te{MeV}$ given in \cite{Amsler:2008zzb} one gets the branching ratios
\begin{align}
\label{eq:branch-om-mu}
 \Gamma_{\omega \rightarrow \pi^0 \mu^+ \mu^-} \ / \ \Gamma_\omega &= (1.16 \pm 0.07) \cdot 10^{-4}, \\
 \Gamma_{\omega \rightarrow \pi^0 e^+ e^-} \ / \ \Gamma_\omega &= (8.1 \pm 0.1) \cdot 10^{-4}
\end{align}
and the experimental ratios given in \cite{Amsler:2008zzb}
\begin{align}
 \Gamma_{\omega \rightarrow \pi^0 \mu^+ \mu^-}^\te{ exp} \ / \  \Gamma_\omega &= (9.6 \pm 2.3) \cdot 10^{-5}, \\
 \Gamma_{\omega \rightarrow \pi^0 e^+ e^-}^\te{ exp} \ / \ \Gamma_\omega &= (7.7 \pm 0.6) \cdot 10^{-4}.
\end{align} 
In order to compare the sensitivity of experiments with muon pairs and with electrons pairs the single-differential decay width of the decay of the $\omega$-meson into a pion and a dielectron is also integrated from $m_{e^+e^-} = 2m_\mu$ on resulting in
\begin{align}
 \Gamma_{\omega \rightarrow \pi^0 e^+ e^-}^\te{ part} &= (1.15 \pm 0.06) \cdot 10^{-6} \ \te{GeV}
\end{align}
and the branching ratio
\begin{align}
\label{eq:branch-om-el}
 \Gamma_{\omega \rightarrow \pi^0 e^+ e^-}^\te{ part} \ / \  \Gamma_\omega &= (1.35 \pm 0.07) \cdot 10^{-4}.  
\end{align}
Obviously, the numbers given in \eqref{eq:branch-om-mu} and \eqref{eq:branch-om-el} are of comparable size
as one can already anticipate from comparing Figs.\ \ref{fig:single-diff decay omega pi mu} and 
\ref{fig:single-diff decay omega pi e teil}.


\section{Decay $\omega \rightarrow \eta \, l^+ l^-$} 
\label{sec:omega eta}

\begin{figure}[h]
 \centering
 \includegraphics[keepaspectratio,width=0.35\textwidth]{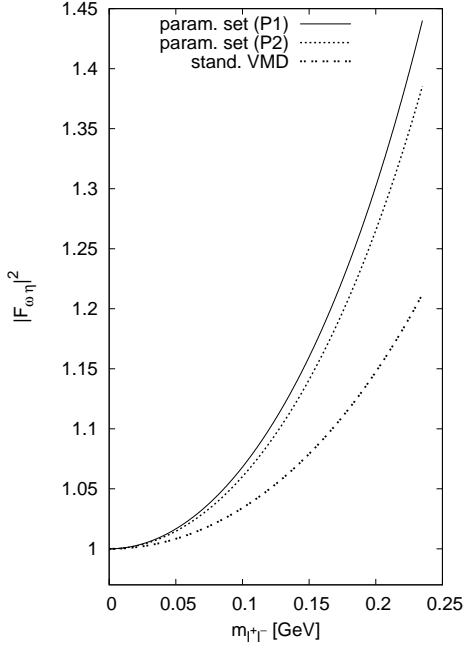}
 \caption{Form factor of the decay $\omega \rightarrow \eta \, l^+ l^-$. The solid line describes the form factor calculated with parameter set \eqref{Ps1} and the dotted the one calculated with parameter set \eqref{Ps2}. The dot-dashed line is calculated with the VMD model.}
 \label{fig:Ff omega eta}
\end{figure}

\begin{figure}[h]
 \centering
 \includegraphics[angle=-90,keepaspectratio,width=0.45\textwidth]{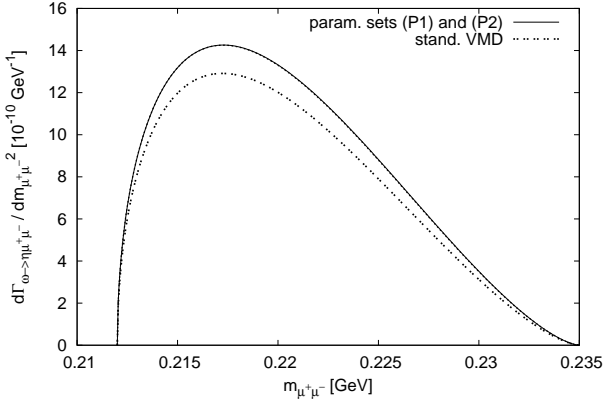}
 \caption{Single-differential decay width of the decay $\omega \rightarrow \eta \, \mu^+ \mu^-$. The solid line describes the (practically indistinguishable) widths calculated with parameter sets \eqref{Ps1} and \eqref{Ps2}, respectively. The dot-dashed line is calculated with the VMD model.}
 \label{fig:single-diff decay omega eta mu}
\end{figure}

\begin{figure}[h]
 \centering
 \includegraphics[angle=-90,keepaspectratio,width=0.45\textwidth]{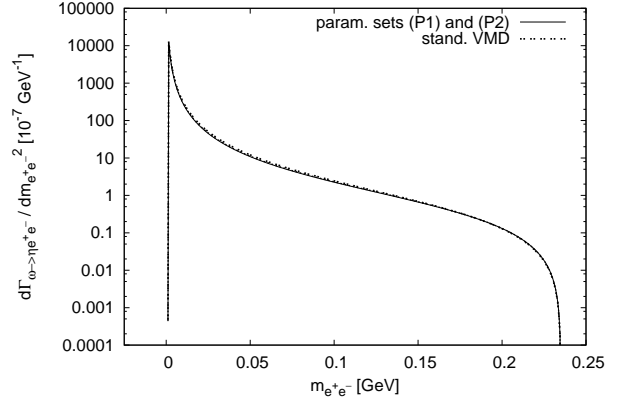}
 \caption{Same as Fig.\ \ref{fig:single-diff decay omega eta mu} but for electrons instead of muons. As they are practically on top of each other the solid line describes the widths calculated with parameter sets \eqref{Ps1}, \eqref{Ps2} and with the VMD model, respectively.}
 \label{fig:single-diff decay omega eta e}
\end{figure}

\begin{figure}[h]
 \centering
 \includegraphics[angle=-90,keepaspectratio,width=0.45\textwidth]{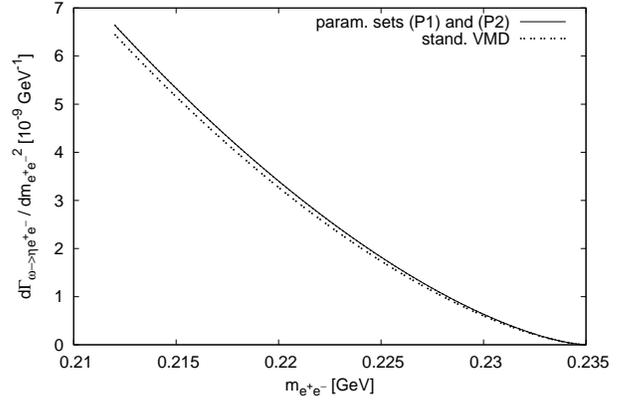}
 \caption{Single-differential decay width of the decay $\omega \rightarrow \eta \, e^+ e^-$ above $m_{e^+e^-} = 2m_\mu$. The again practically indistinguishable widths calculated with the parameter sets \eqref{Ps1} and \eqref{Ps2} are described by the solid line, the widths calculated with the VMD model is described by the dot-dashed line.}
 \label{fig:single-diff decay omega eta e teil}
\end{figure}

From the Lagrangian \eqref{eq:Lid} one deduces that the indirect decay of the $\omega$-meson into an 
$\eta$-meson and a dilepton can only happen via a virtual $\omega$-meson. 
The form factor including both the indirect and the direct decay is
\begin{align}
 f_{\omega\eta}(q^2) =& \frac{m_\omega}{6 \sqrt{3} f m_V e} \left[ e_A + 2 b_A e_V m_V^2 \frac{m_\pi^2}{m_\omega^2} S_\omega(q^2) \right. \\ \nn 
 &\left. - \frac{1}{4} e_V h_A m_V^2 \left(1+\frac{q^2}{m_\omega^2}\right)S_\omega(q^2) \right]
\end{align}
with the $\omega$-meson propagator defined analogously to the $\rho$-meson propagator in Eq.\ \eqref{eq:rho propagator}. The form factor is plotted in Fig.\ \ref{fig:Ff omega eta} and the single-differential decay widths for the decays $\omega \rightarrow \eta \, \mu^+ \mu^-$ and $\omega \rightarrow \eta \, e^+ e^-$ 
in Figs.\ \ref{fig:single-diff decay omega eta mu} and \ref{fig:single-diff decay omega eta e}, respectively. The single-differential decay width for the decay $\omega \rightarrow \eta \  e^+ e^-$ above $m_{e^+e^-} = 2m_\mu$ is plotted in Fig.\ \ref{fig:single-diff decay omega eta e teil}. As in the 
previously discussed reaction the results obtained from our two parameter sets do not deviate much from each other.
Whereas the result for the decay width calculated with the VMD model is off our results for a decay into a dimuon, all three curves are consistent with each others for the case of a decay into a dielectron.

As there are no experimental data for these decays available, the following partial decay widths are predictions. With our Lagrangian one gets for the partial decay widths
\begin{align}
 \Gamma_{\omega \rightarrow \eta \, \mu^+ \mu^-} &= (8.51 \pm 0.01) \cd 10^{-12} \,\te{GeV}, \\
 \Gamma_{\omega \rightarrow \eta \, e^+ e^-} &= (2.72 \pm 0.09) \cd 10^{-8} \,\te{GeV}
\end{align}
and the branching ratios
\begin{align}
 \Gamma_{\omega \rightarrow \eta \, \mu^+ \mu^-} \ / \  \Gamma_\omega &= (1.00 \pm 0.00) \cdot 10^{-9}, \\
 \Gamma_{\omega \rightarrow \eta \, e+ e-} \ / \  \Gamma_\omega &= (3.20 \pm 0.10) \cdot 10^{-6}.
\end{align}
In view of the order of magnitude of the partial decay width and the branching ratio for the decay into a dimuon, a verification of this result by experiments might not be possible.\\
Additionally, one gets the partly integrated single-differential decay width of the decay into a dielectron 
\begin{align}
 \Gamma_{\omega \rightarrow \eta \ e^+ e^-}^\te{ part} = (2.61 \pm 0.00) \cdot 10^{-11} \ \te{GeV}
\end{align}
and the corresponding branching ratio
\begin{align}
 \Gamma_{\omega \rightarrow \eta \ e^+ e^-}^\te{ part} \ / \  \Gamma_\omega = (3.07 \pm 0.00) \cdot 10^{-9}.
\end{align}


\section{Decay $\phi \rightarrow \eta \, l^+ l^-$} 
\label{sec:phi eta}

\begin{figure}[h]
 \centering
 \includegraphics[keepaspectratio,width=0.35\textwidth]{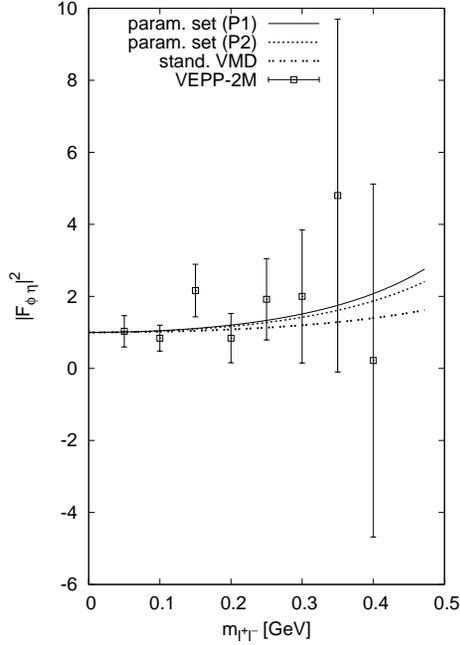}
 \caption{Form factor of the decay $\phi \rightarrow \eta \, l^+ l^-$ compared to the experimental data (for electrons) taken at the VEPP-2M collider \cite{Achasov:2000ne}. The solid line describes the form factor calculated with parameter set \eqref{Ps1} and the dotted the one calculated with parameter set \eqref{Ps2}. The dot-dashed line is obtained with the VMD model.}
 \label{fig:Ff phi eta}
\end{figure}

The Lagrangian \eqref{eq:Lid} only allows for an indirect decay of the $\phi$-meson into an $\eta$-meson and a dilepton 
via a virtual $\phi$-meson. The form factor including both the indirect and the direct decay equals

\begin{align}
 f_{\phi\eta}(q^2) =& \frac{2 m_\phi}{3 \sqrt{6} f m_V e} \left[ e_A + 2 b_A e_V m_V^2 \frac{2m_k^2-m_\pi^2}{m_\phi^2} S_\phi(q^2) \right. \\ \nn 
 &\left. - \frac{1}{4} e_V h_A m_V^2 \left(1+\frac{q^2}{m_\phi^2}\right)S_\phi(q^2) \right]
\end{align}
with the $\phi$-meson propagator defined analogously to the $\rho$-meson propagator in Eq.\ \eqref{eq:rho propagator}. In Fig.\ \ref{fig:Ff phi eta} the form factor is plotted in comparison to the data taken with the SND detector at the VEPP-2M collider \cite{Achasov:2000ne} for the decay of a $\phi$-meson into an $\eta$-meson and a dielectron. Although our calculations are in agreement with the data, it is not possible to evaluate how well they describe the data due to the relatively large error bars.
Also the VMD model agrees with the data within errors. Deviations between the results obtained from our two parameter sets are
small. 
Better data can help to see whether our approach or the standard VMD model does a better job. 

\begin{figure}[h]
 \centering
 \includegraphics[angle=-90,keepaspectratio,width=0.45\textwidth]{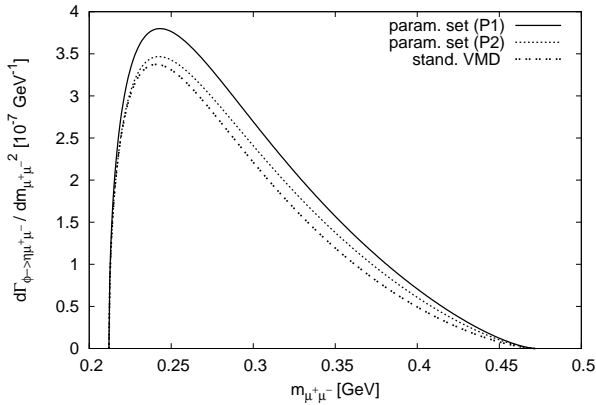}
 \caption{Single-differential decay width of the decay $\phi \rightarrow \eta \, \mu^+ \mu^-$. 
The solid/dotted line describes the width calculated with parameter set \eqref{Ps1}/\eqref{Ps2}. 
The dot-dashed line is obtained with the VMD model.}
 \label{fig:single-diff decay phi eta mu}
\end{figure}

\begin{figure}[h]
 \centering
 \includegraphics[angle=-90,keepaspectratio,width=0.45\textwidth]{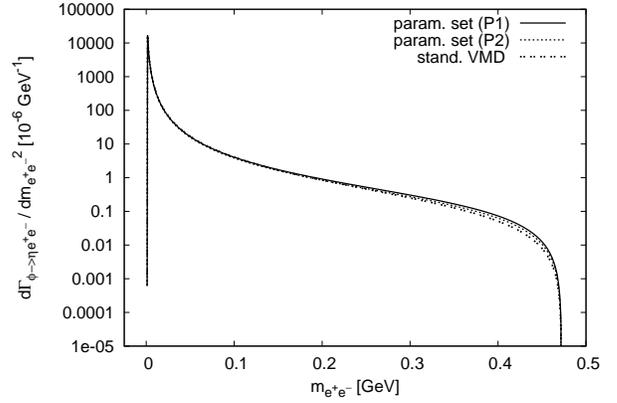}
 \caption{Same as Fig.\ \ref{fig:single-diff decay phi eta mu} but for electrons instead of muons.}
 \label{fig:single-diff decay phi eta e}
\end{figure}

\begin{figure}[h]
 \centering
 \includegraphics[angle=-90,keepaspectratio,width=0.45\textwidth]{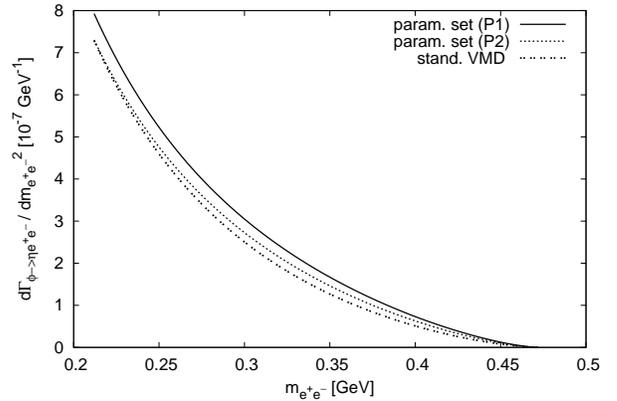}
 \caption{Same as Fig.\ \ref{fig:single-diff decay phi eta e} but only for dielectron masses above $2m_\mu$.}
 \label{fig:single-diff decay phi eta e teil}
\end{figure}

In Figs.\ \ref{fig:single-diff decay phi eta mu} and \ref{fig:single-diff decay phi eta e}, \ref{fig:single-diff decay phi eta e teil} the single-differential decay widths for the decays $\phi \rightarrow \eta \, \mu^+ \mu^-$ and $\phi \rightarrow \eta \, e^+ e^-$, respectively, are plotted.

Again, the three different theoretical single-differential decay widths for the decay $\phi \rightarrow \eta \, e^+ e^-$ agree well. For the decay $\phi \rightarrow \eta \, \mu^+ \mu^-$ the differences between the curves calculated with parameter set \eqref{Ps1} and \eqref{Ps2} are larger than those for all other decays. 
This might seem to be in contradiction to Fig.\ \ref{fig:Ff phi eta} where the two corresponding curves were
closer together. We stress, however, that we do not use the experimental value of $\Gamma_{\phi \to \eta \gamma}$
to get the differential width from \eqref{eq:single-diff} but rather the respective one determined in 
\eqref{eq:BreiteAgamma}.
By inspecting Table \ref{tab:Parameter} one sees that the corresponding values for our two parameter sets deviate
from each other. This has the effect to move the \eqref{Ps2} curve downwards and away from the \eqref{Ps1} curve.
Remembering that on the one hand the $\phi$-meson is the heaviest light vector meson and on the other hand the differences between the parameter sets roughly give the error of the leading-order calculation, this is an intrinsically consistent result of our approach.

For the partial decay widths one gets
\begin{align}
 \Gamma_{\phi \rightarrow \eta \, \mu^+ \mu^-} &= (2.75 \pm 0.29) \cd 10^{-8} \,\te{GeV}, \\
 \Gamma_{\phi \rightarrow \eta \, e^+ e^-} &= (4.64 \pm 0.26) \cd 10^{-7} \,\te{GeV}.
\end{align}
and with the full width $\Gamma_\phi = (4.26 \pm 0.04) \ \te{MeV}$ taken from \cite{Amsler:2008zzb} one gets for the branching ratios
\begin{align}
 \Gamma_{\phi \rightarrow \eta \ \mu^+ \mu^-} \ / \  \Gamma_\phi &= (6.44 \pm 0.69) \cdot 10^{-6}, \\
 \Gamma_{\phi \rightarrow \eta \ e+ e-} \ / \  \Gamma_\phi &= (1.09 \pm 0.06) \cdot 10^{-4}.
\end{align}
For the decay into a dimuon no experimental values are available. The calculated values for the decay into a dielectron agree well with the experimental values given in \cite{Amsler:2008zzb}
\begin{align}
 \Gamma_{\phi \rightarrow \eta \, e^+ e^-}^{\te{ exp}} = (4.90 \pm 0.47) \cd 10^{-7} \,\te{GeV}
\end{align}
and 
\begin{align}
 \Gamma_{\phi \rightarrow \eta \ e^+ e^-}^\te{ exp} \ / \  \Gamma_\phi = (1.15 \pm 0.10) \cdot 10^{-4}.
\end{align}
The value for the partly integrated single-differential decay width of the decay into a dielectron is
\begin{align}
 \Gamma_{\phi \rightarrow \eta \ e^+ e^-}^\te{ part} = (3.59 \pm 0.37) \cd 10^{-8} \ \te{GeV}
\end{align}
with the branching ratio
\begin{align}
 \Gamma_{\phi \rightarrow \eta \ e^+ e^-}^\te{ part} \ / \  \Gamma_\phi = (8.43 \pm 0.87) \cdot 10^{-6}.
\end{align}


\section{Summary and outlook} \label{sec:summary}

The chiral Lagrangian including light vector mesons and Goldstone bosons was used to calculate in leading order the 
decays of narrow light vector mesons into a pseudoscalar meson and a dilepton. Thereby, the leading-order terms were identified by the counting rules proposed in \cite{Lutz:2008km}. In general, the results are in very good agreement with the available experimental data. We predict several quantities not determined yet by experiments. Concerning the form factor of the
$\omega$ to  $\pi^0$ transition we have obtained a much better description of the NA60 data than the standard VMD model. 
Only for high dilepton masses close to the kinematical boundary we failed to describe the very steep rise of the data. 
In view of the fact that so far no satisfying theoretical description of the large deviations from VMD was available
(cf.\ the discussions in \cite{Klingl:1996by,Faessler:1999de,Friman:1995qm}), we regard our approach as an important step forward. 

In the present work we have restricted ourselves more or less to a leading-order calculation. We have tried to estimate
the error induced by that restriction by keeping one (tree-level) next-to-leading-order term. The deviations turned out to be small. Nonetheless, to show that our whole approach makes sense
as an effective field theory and not just as a cleverly chosen hadronic tree-level model it is mandatory to perform at least
next-to-leading-order calculations. Also in that context we expect that interesting interrelations will show up by a combined
study of reactions like $\omega \to \eta \gamma^*$ and $\eta \to \gamma^{(*)} \omega^* \to \gamma^{(*)} \gamma^*$. 
It might appear that in a next-to-leading-order calculation, being valid up to higher energies, one also gets a better description of the high-mass part of the $\omega$ to $\pi^0$ transition form factor. 

Logically prior to the next-to-leading-order calculations is an exploratory tree-level calculation of the mentioned 
pseudoscalar decays. Here
further experimental results for all available channels like $\pi,\eta$ to $\gamma$ + dilepton and also to
dilepton + dilepton would be extremely helpful.

\section*{Acknowledgments}

We acknowledge encouragement and countless suggestions how to improve the manuscript from A.\ Kup{\'s}{\'c}.
We thank S.\ Damjanovic for providing us with data from NA60. 
We also thank her and V.\ Metag, M.\ Soyeur and J.\ Specht for very helpful discussions 
and comments on the manuscript.
This work has been supported by GSI Darmstadt. 



\bibliography{literature}
\bibliographystyle{apsrev4-1}

\end{document}